\documentclass[%
reprint,
superscriptaddress,
preprintnumbers,
amsmath,amssymb,
aps,
prd,
]{revtex4-2}

\usepackage[utf8]{inputenc}

\usepackage{mathtools}
\usepackage{amsfonts}
\usepackage{mathrsfs}
\usepackage{bm}
\usepackage{bbold}
\usepackage{slashed}
\usepackage{dsfont}
\usepackage{float}
\usepackage{dcolumn}
\usepackage{amssymb,amsmath}

\usepackage{graphicx}
\usepackage{subcaption}
\usepackage{color}
\usepackage{array}

\usepackage{placeins}
\usepackage{booktabs}
\usepackage{caption}

\usepackage{xspace}
\usepackage{hyperref}
\usepackage[nameinlink]{cleveref}
\usepackage{bookmark}
\usepackage{units}

\def\Eq#1{Eq.~\labelcref{#1}}

\def\Fig#1{Fig.~\labelcref{#1}}

\def\sec#1{Sec.~\labelcref{#1}}
\def\app#1{App.~\labelcref{#1}}

\setkeys{Gin}{width=0.48\textwidth}
\usepackage{booktabs}
\usepackage{multirow}

\newcolumntype{C}{>{$}c<{$}}
\AtBeginDocument{
	\heavyrulewidth=.08em
	\lightrulewidth=.05em
	\cmidrulewidth=.03em
	\belowrulesep=.65ex
	\belowbottomsep=0pt
	\aboverulesep=.4ex
	\abovetopsep=0pt
	\cmidrulesep=\doublerulesep
	\cmidrulekern=.5em
	\defaultaddspace=.5em
}

\graphicspath{{./figures/}}

\usepackage{xifthen}
\usepackage{xcolor}

\newcommand{\gettitle}{Critical dynamics of Model H within the real-time fRG approach}

\hypersetup{
	colorlinks,
	linkcolor={red!75!black},
	citecolor={blue!75!black},
	urlcolor={blue!75!black}, 
	pdftitle={\gettitle},
	pdfauthor={},
	pdfkeywords={}
	{} 
	bookmarksopen=true,
	bookmarksopenlevel=2,
	bookmarksnumbered=true
}

\begin{document}
\title{\gettitle}

\author{Yong-rui Chen}
\affiliation{School of Physics, Dalian University of Technology, Dalian, 116024, P.R. China}	

\author{Yang-yang Tan}
\affiliation{School of Physics, Dalian University of Technology, Dalian, 116024, P.R. China}

\author{Wei-jie Fu}
\affiliation{School of Physics, Dalian University of Technology, Dalian, 116024, P.R. China}

\begin{abstract}

The critical dynamics of Model H with a conserved order parameter coupled to a transverse momentum density which describes the gas-liquid or binary-fluid transitions is investigated within the functional renormalization group approach formulated on the closed time path. According to the dynamic scaling analysis, Model H and QCD critical end point belong to the same dynamic universality class in the critical region. The higher-order correction of the transport coefficient $\bar\lambda$ and shear viscosity $\bar\eta$ arising from mode-couplings are obtained by calculating the two-point correlation functions. The flow equation of a dimensionless coupling constant for nondissipative interactions is derived to look for the fixed-point solution of the system. The scaling relation between the critical exponent of the transport coefficient and that of the shear viscosity is estimated. Finally, the dynamic critical exponent $z$ is obtained as a function of the spatial dimension $d$.

\end{abstract}

\maketitle

\section{Introduction}
\label{sec:int}

Critical dynamics plays a crucial role in the studies of QCD phase structure that is related to the phase transition from the confinement hadronic phase to the deconfinement phase of quark-gluon plasma (QGP) \cite{Fischer:2018sdj, Dupuis:2020fhh, Fu:2022gou, Braun:2023qak}. It is expected that fluctuations of conserved charges can be used to search for the critical end point (CEP) in heavy-ion collision experiments \cite{Bluhm:2020mpc, Luo:2017faz, Stephanov:2008qz, Fu:2021oaw, Fu:2023lcm}. Significant progresses have been made in the Beam Energy Scan (BES) program at Relativistic Heavy Ion Collider (RHIC) \cite{STAR:2013gus, STAR:2017sal, Bzdak:2019pkr, STAR:2020tga}. Notably, recent estimates of the location of CEP in the QCD phase diagram have arrived at convergent results from functional QCD \cite{Fu:2019hdw, Gao:2020fbl, Gunkel:2021oya}, which are also consistent with the results obtained from finite-size scaling of proton cumulants in heavy-ion collisions \cite{Sorensen:2024mry} and from extrapolation of lattice simulations \cite{Clarke:2024ugt}. The correlation length tends to be divergent near the CEP, such that properties of the critical behavior are universal and independent of the detail of microscopic interactions. It is found that the chiral phase transition at vanishing chemical potential in the chiral limit of $u$ and $d$ quarks is classified into the universality of three dimensional ($3d$) $O(4)$ symmetry \cite{Pisarski:1983ms}, and the CEP in the QCD phase diagram belongs to the $Z(2)$ universality class \cite{Halasz:1998qr}, see also e.g., \cite{Chen:2021iuo} for relevant discussions.

The criticality is not only reflected in static equilibrium properties, but also in real-time nonequilibrium dynamics. When the critical end point is approached, the relaxation time from nonequilibrium to equilibrium is significantly increased, that is known as the effect of critical slowing down \cite{Rajagopal:1992qz, Berdnikov:1999ph}. The renormalization group technique by Wilson \textit{et al}. have been extended to study the dynamic critical phenomena \cite{Halperin:1972bwo,  Halperin:1974zz, Halperin:1976zz, Halperin:1976zza}, where several characteristic dynamic universalities were investigated and classified, see \cite{Hohenberg:1977ym, Folk_2006} for related reviews. It has been shown that the second-order chiral phase transition of QCD lies in the dynamic universality class of `Model G' \cite{Rajagopal:1992qz}, and the dynamics of QCD critical end point belongs to `Model H' \cite{Son:2004iv}. A plethora of results about the real-time correlation functions and critical dynamics have been obtained within the classical-statistical lattice simulations \cite{Berges:2009jz, Schlichting:2019tbr, Schweitzer:2020noq, Schweitzer:2021iqk, Florio:2021jlx, Florio:2023kmy}. Recently, the functional renormalization group (fRG) within the Schwinger-Keldysh formalism \cite{Berges:2008sr, Gasenzer:2007za, Pawlowski:2015mia, Corell:2019jxh, Braun:2022mgx} has been used to investigate the spectral functions \cite{Huelsmann:2020xcy, Tan:2021zid, Roth:2021nrd}, pseudo-Goldstone damping \cite{Tan:2024fuq}, critical dynamics and dynamic critical exponents in Model A \cite{Roth:2023wbp, Chen:2023tqc, Canet:2006xu, Canet:2011wf, Batini:2023nan}, Model C \cite{Roth:2023wbp, Mesterhazy:2013naa} and Model G \cite{Roth:2024rbi} and so forth. This provides us with a powerful tool to explore the critical dynamics of fluctuations with nonperturbative interactions.

In this work we would like to study the properties of dynamic universality class `Model H' using the real-time fRG approach. Model H is characterized by a conserved order parameter field, which can be treated as a combination of the chiral condensate and baryon density in QCD, and this order parameter couples with the energy-momentum density. That is related to the liquid-gas phase transition. Previous studies focus on the perturbative $\epsilon$-expansion method, see \cite{Siggia:1976zz, Yee:2017sir}. Recently stochastic fluid dynamics simulation has also been performed on Model H \cite{Chattopadhyay:2024jlh}. From the equation of motion, we construct the effective action in terms of two different kinds of field with the Keldysh rotation \cite{Schwinger:1960qe, Keldysh:1964ud}. The mode-couplings and dissipative dynamics are encoded through the calculations of correlation functions. By solving the fixed-point equations we obtain the anomalous dimensions and dynamic critical exponent.
 
This paper is organized as follows: In \sec{sec:langevin} we briefly introduce the Langevin equation of Model H. The formalism of Model H formulated on the closed time path within the fRG approach is specialized in \sec{sec:ModelH}, where the flow equations of the effective potential and relevant parameters are shown. Numerical results are presented and discussed in \sec{sec:num-resul}. In \sec{sec:summary} we give a summary and conclusion. The expressions of the propagators and Feynman rules of vertices are collected in \app{app:pro_vertex}. In \app{app:loop}, we show a specific calculation for loop diagram.

\section{Langevin equation for Model H}
\label{sec:langevin}
The simplified model that describes the dynamics of the gas-liquid or binary-fluid transition was introduced earlier in the mode-coupling theories, which was denoted as Model H \cite{Siggia:1976zz, Hohenberg:1977ym}. The equations of motion for the coupled modes in Model H are given by the Langevin equations
\begin{align}
    \frac{\partial\phi}{\partial t}&=\lambda_0\bm{\nabla}^2 \frac{\delta\mathscr{H}}{\delta\phi}-g_0\bm{\nabla} \phi\cdot\frac{\delta\mathscr{H}}{\delta\bm{j}}+\theta(x,t)\,,\label{eq:langEq-phi}\\
    \frac{\partial\bm{j}}{\partial t}&=\Pi^{\bot} \left(\eta_0 \bm{\nabla}^2 \frac{\delta\mathscr{H}}{\delta\bm{j}} + g_0 \bm{\nabla}\phi \frac{\delta\mathscr{H}}{\delta\phi}
    +\bm{\zeta}(x,t)\right)\,,\label{eq:langEq-j}
\end{align}
with the Hamiltonian
\begin{align}
    \mathscr{H}=\int \mathrm{d}^d x\left(\frac12\vert\bm{\nabla}\phi(x)\vert^2 +\frac{r}{2}\phi^2 +\frac{u}{4}\phi^4 + \frac12\bm{j}^2\right)\,.
\end{align}
The correlations of Gaussian noises read
\begin{align}
    \langle \theta(x,t) \theta(x',t')\rangle=&-2\lambda_0\bm{\nabla}^2\delta(x-x')\delta(t-t'),\nonumber\\
    \langle \zeta_\mu(x,t) \zeta_\nu(x',t')\rangle=&-2\eta_0\bm{\nabla}^2\delta(x-x')\delta(t-t')\delta_{\mu\nu}.\label{eq:noise_cor}
\end{align}
The scalar field $\phi$ of order parameter is a conserved quantity. The vector field $\bm{j}$ corresponds to the momentum density or velocity, that has only the transverse components, since the right hand side of equation of motion for $\bm{j}$ in \Eq{eq:langEq-j} is multiplied by the transverse projection operator, i.e.,
\begin{align}
    \Pi^\bot_{\alpha \beta}=&(\delta_{\alpha\beta}-\bm{q}_\alpha \bm{q}_\beta/\bm{q}^2)\,.\label{}
\end{align}
In \Eq{eq:langEq-phi,eq:langEq-j} $\lambda_0$ and $\eta_0$ are the transport coefficient and shear viscosity, respectively. Their dynamical scaling behaviors are described by the exponents $x_\lambda$ and $x_\eta$ in the critical regime, and these exponents can also be seen as dynamic anomalous dimensions. The Hamiltonian has the usual form except the term $\bm{j}^2/2$ included. Thus the static critical properties of $\phi$ are the same as the Ginzburg-Landau model with Ising symmetry. When the system is in thermal equilibrium, the equilibrium distribution function is given by $e^{-\mathscr{H}}$. The coupling constant $g_0$ describes the interaction between the two dynamical modes in model H. The Gaussian noises $\theta$ and $\zeta$ drive the system to thermal equilibrium and their correlation functions are shown in \Eq{eq:noise_cor}.
 
According to the dynamic scaling relation, the characteristic frequency of the order parameter has the following form \cite{Halperin:1969rhg}
\begin{align}
    \omega_\phi(k)=Dk^2\Omega(k\xi)=D_0\xi^{2-z}k^2\Omega(k\xi)\,,\label{}
\end{align}
in the limit $\xi\to\infty,\,k\to 0$. Here $\xi$ is the correlation length, and the diffusion constant $D$ satisfies the `Kawasaki-Stokes' relation \cite{Arcovito:1969, Lo:1973}
\begin{align}
    D\equiv\frac{\lambda}{\chi_\phi}=\frac{R\,k_BT}{\eta \xi} \,,\label{eq:KS-relation}
\end{align}
in $d=3$ dimensions. Here $\lambda$ and $\eta$ stand for the renormalized transport coefficient and shear viscosity, respectively; $\chi_\phi$ denotes the static susceptibility; $T$ is the temperature and $k_B$ the Boltzmann constant; $R$ is a constant. Using the scaling relations as follows
\begin{align}
    \chi_\phi\sim \xi^{2-\eta_\phi}, \quad \lambda \sim \xi^{x_\lambda}, \quad \eta \sim \xi^{x_\eta}\,,\label{eq:scaling}
\end{align}
one arrives at the dynamic critical exponent $z$ for the order parameter
\begin{align}
    z=4-\eta_\phi-x_\lambda\,, \label{eq:exponentz}
\end{align}
where $\eta_\phi$ is the static anomalous dimension. Note that the Kawasaki-Stokes relation in \Eq{eq:KS-relation} also leaves us with
\begin{align}
   x_\eta+x_\lambda=1-\eta_\phi\,, \label{}
\end{align}
in $d=3$ dimensions. The  results of the exponents $x_\lambda$ and $x_\eta$ in the $\epsilon$-expansion up to the order of $O(\epsilon^2)$ are obtained as \cite{Siggia:1976zz}
\begin{align}
    x_\lambda=\frac{18}{19}\epsilon[1-0.033\epsilon+O(\epsilon^2)]\,,\nonumber\\
    x_\eta=\frac{1}{19}\epsilon[1+0.238\epsilon+O(\epsilon^2)]. \label{eq:xlam-eta-expan}
\end{align}
In the next section, we will compute the aforementioned exponents using the functional renormalization group approach.

\section{Model H within the real-time fRG approach}
\label{sec:ModelH}
%
%

The crucial part of the functional renormalization group (fRG) approach is the scale-dependent effective action $\Gamma_k[\Phi]$ that interpolates smoothly between the short-distance and long-distance physics with the evolution of the exact renormalization group equation \cite{Wetterich:1992yh}, for fRG reviews see \cite{Pawlowski:2005xe,Dupuis:2020fhh,Fu:2022gou}. At the ultraviolet cutoff $k=\Lambda$, the effective action
$\Gamma_\Lambda[\Phi]$ coincides with the classical action $S[\Phi]$ and at $k=0$ the effective action is the quantum action as fluctuations of different momentum modes are integrated in successively. To formulate the effective action in the closed time path, starting from the Langevin equation in \Eq{eq:langEq-phi,eq:langEq-j}, we perform the path integral about the noise variables that result in the Martin-Siggia-Rose response field \cite{Martin:1973zz}. The renormalization group (RG) scale $k$ dependent effective action of Model H in the Schwinger-Keldysh formalism is
\begin{align}
    \Gamma_k[\Phi]&=i\int\mathrm{d}^d x\mathrm{d}t \,\,\Bigg\{\phi_q\left(\frac{\partial}{\partial t}\phi_c+\lambda_k \bm{\nabla}^2 \bm{\nabla}^2\phi_c\right)
    \nonumber\\&-\lambda_k\phi_q \bm{\nabla}^2 \frac{\delta V_k(\rho_c)}{\delta \phi_c} 
    +g_k\phi_q \bm{\nabla}\phi_c \cdot\bm{j_c}
    +2\lambda_k \phi_q\bm{\nabla}^2\phi_q \nonumber\\    &+\bm{j}_{\bm{q},\alpha}\Pi^\bot_{\alpha\beta}\Bigg[ \left(\frac{\partial}{\partial t}\bm{j}_{\bm{c},\beta}
    -\eta_k \bm{\nabla}^2 \bm{j}_{\bm{c},\beta}\right)
    \nonumber\\&-g_k\bm{\nabla}\phi_c(-\bm{\nabla}^2\phi_c)+2\eta_k \bm{\nabla}^2\bm{j}_{\bm{q},\beta}\Bigg]\Bigg\}\,.\label{eq:Gamk}
\end{align}
Here, $(\phi_c, \phi_q)$ and $(\bm{j}_{\bm{c},\beta}, \bm{j}_{\bm{q},\alpha})$ are the order parameter fields and momentum densities with the subscripts `c' and `q' denoting classical and quantum fields, respectively, see \cite{Tan:2021zid, Chen:2023tqc} for more details. And subscripts $\alpha,\beta$ denote the component of vector $\bm{j}$, i.e., $\alpha,\beta=1,2,...,d$. Different from the analogue in \Eq{eq:langEq-phi,eq:langEq-j}, the transport and coupling constants $\lambda_k,\eta_k$ and $g_k$ in \Eq{eq:Gamk} also receive corrections from interactions, and they are RG-scale dependent. Note that the order parameter belongs to the Ising universality class, thus presents the same static critical properties. The order parameter can be regarded as a linear combination of the chiral condensate and baryon density in QCD \cite{Son:2004iv}. Moreover, one can see that the powers of gradient in front of $\phi_c$ are higher than those in Model A or Model C due to the conservation of the order parameter and momentum. The higher-order scatterings among the $\phi$ fields as well as the mass term of order parameter are encoded in the effective potential. 

\begin{figure}[htbp]
\includegraphics[width=0.45\textwidth]{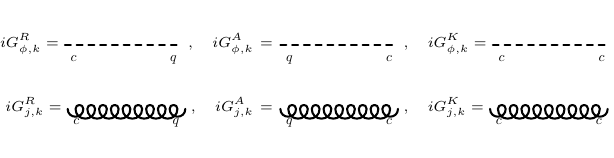}
\caption{Diagrammatic representation of the retarded, advanced and Keldysh propagators for the order parameter  and momentum density fields in this work.}\label{fig:mesonprop}
\end{figure}

The flow equation of the effective action reads  
\begin{align}
    \partial_\tau\Gamma_k[\Phi]=\int G_k\partial_\tau R_k \,,\label{eq:wetterichequ}
\end{align}
with the propagator $G_k$
\begin{align}
    G_k=\frac{1}{\Gamma_k^{(2)}[\Phi]+R_k} \,,\label{}
\end{align}
where $R_k$ stands for the regulator and $\Gamma_k^{(2)}[\Phi]$ the second derivative of $\Gamma_k$ with respect to the fields. In this context, the related fields are the order parameter and transverse momentum density. We present the expression of the propagators and three-point coupling vertices in \app{app:pro_vertex}. To be simplified, we directly show the matrix form of the propagator that is
\begin{align}
  G_{k}=&\begin{pmatrix} G^{K}_{k} &G^R_k\\[1ex] G^{A}_{k}  &0 \end{pmatrix}\,,\label{eq:prop}
\end{align}
where $K,R$ and $A$ denote the Keldysh, retarded and advanced propagators, respectively. In \Fig{fig:mesonprop} we show the diagrammatic representation for the relevant propagators, the dashed lines denote the order parameter and the wave lines denote the vector momentum density.  
%
%

%
\begin{figure}[t]
\includegraphics[width=0.4\textwidth]{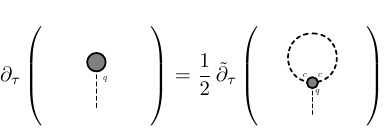}
\caption{Diagrammatic representation of the flow equation for the effective potential obtained from the one-point correlation function of the effective action actually. The external leg denotes the field $\phi_q$ and the internal line denotes the Keldysh type propagator for the order parameter.}\label{fig:Gam1-phiq-equ}
\end{figure}
%
%
\begin{figure}[t]
\includegraphics[width=0.5\textwidth]{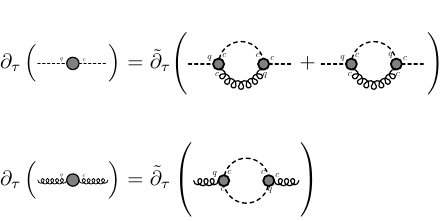}
\caption{Diagrammatic representation of the flow equation for the inverse retarded propagator related to the order parameter and momentum density fields, see \Eq{eq:dtlameta}.}\label{fig:Gam2-equ}
\end{figure}
%

In the following, we will apply the real-time fRG approach to compute the flow equations of effective potential $V_k(\rho)$, transport coefficient $\lambda_k$ and shear viscosity $\eta_k$ and coupling constant $g_k$. The flow of effective potential corresponds to the one-point correlation function of $\Gamma_k[\Phi]$ with respect to the field $\phi_q$ as shown in \Fig{fig:Gam1-phiq-equ},
\begin{align}
    \partial_\tau V'_k(\rho)=\int_{\omega,q}\frac12 \Gamma^{(3)}_{\phi_q\phi_c\phi_c,k} \Tilde{\partial}_\tau G^K_{\phi\phi,k}/(-\lambda_k \bm{\nabla}^2)\,, \label{eq:flowV}
\end{align}
with the explicit expression of vertex $\Gamma^{(3)}_{\phi_q\phi_c\phi_c,k}$ shown in \app{app:pro_vertex}. The flow equation of $V'_k(\rho)$ is the same as Model A with one component of the order parameter.

The flow for the $\lambda_k$ and $\eta_k$ are related to the inverse of retarded propagators for the order parameter and momentum density fields respectively, which can be obtained by projecting onto the external momentum direction, to wit,
\begin{align}
  \partial_{\tau} \lambda_k=&\lim_{\substack{p_0\rightarrow 0\\ \bm{p}\rightarrow 0}}(-i)\frac{\partial}{\partial\bm{p}^4}\frac{\delta^2 \partial_\tau\Gamma_{k}[\Phi]}{\delta \phi_{q}(-p)\delta \phi_{c}(p)}\bigg\vert_{\Phi_{\mathrm{EoM}}}\,,\nonumber\\
  \partial_{\tau} \eta_k=&\frac{1}{d-1} \lim_{\substack{p_0\rightarrow 0\\ \bm{p}\rightarrow 0}} (-i)\frac{\partial}{\partial \bm{p}^2}  \frac{\delta^2 \partial_\tau \Gamma_k[\Phi]}{\delta \bm{j}_{\bm{q},\alpha}(-p) 
  \delta\bm{j}_{\bm{c},\beta}(p)} \Pi^\bot_{\alpha\beta}\bigg\vert_{\Phi_{\mathrm{EoM}}}\,,
  \label{eq:dtlameta}
\end{align}
with the factor $1/(d-1)$ in the flow of $\eta_k$ arising from the transverse projection operator. We show the diagrammatic representation for the two-point correlation function of the corresponding fields in \Fig{fig:Gam2-equ}. Here the partial operator $\tilde{\partial}_\tau$ hits the $k$-dependence only through the regulator in propagators. This provides us with all the possible diagrams we need to calculate. And the types of propagators and vertices are labelled in the diagrams with the letters `q' and `c'. Since the external momentum information is contained in the vertices and propagators, the final result on the power of external momenta should account for the contributions from both the vertices and propagators. 

Here we also introduce the dimensionless quantities to convert the flow equations to fixed-point equations which is more convenient to investigate the scaling behavior, such as
\begin{align}
\left \{\begin{array}{l}
   \bar{\lambda}=k^{4-z}\lambda_k\\[1ex]
   \bar{\eta}=k^{2-z}\eta_k\\[1ex]
   \bar{g}=k^{1+\frac{d}{2}-z}g_k\\[1ex]
   \bar{\rho}=k^{2-d}\rho\\[1ex]
   u(\bar{\rho})=k^{-d}V_k(\rho) 
\end{array}\right. \,.\label{eq:dimensionless}
\end{align}

In \app{app:loop}, a specific example for the loop diagram calculation is presented for illustrative purpose, where we also show the expansion up to certain power of external momentum and the related approximation used there. Performing the integrals on the right hand side of \Fig{fig:Gam1-phiq-equ} and \Fig{fig:Gam2-equ}, one arrives at the final form of the flow equations for the effective potential $u'(\bar{\rho})$, and the coefficients $\bar{\lambda},\bar{\eta},  \bar{g}$ as follows
\begin{align}
&\partial_\tau u'(\bar{\rho})\nonumber\\
=&(-2+\eta_\phi)u'(\bar{\rho})+(-2+d+\eta_\phi)\bar{\rho}u^{(2)}(\bar{\rho})\nonumber\\
&-\frac{2\nu_d}{d}
    \left(1-\frac{\eta_\phi}{d+2}\right)\left(\frac{3u^{(2)}(\bar{\rho})+2\bar{\rho}u^{(3)}(\bar{\rho})}{(1+u'(\bar \rho)+2\bar \rho u^{(2)}(\bar \rho))^2}\right)\,,\label{eq:flow_potential}\\
&\partial_\tau \bar{\lambda}\nonumber\\
=&-(z-4+\eta_\phi)\bar{\lambda}\nonumber\\
 &-\frac{2\bar{g}^2}{\bar{\eta}}\frac{1}{(1+\bar{m}^2)^2}\int\mathrm{d\Omega_d} 
 (-1+\cos^2\theta) \cos^2\theta\nonumber\\
&+\frac{2}{d}\frac{2\bar{g}^2}{\bar{\eta}} \frac{1}{(1+\bar{m}^2)^2} \int\mathrm{d\Omega_d}
(2-6\cos^2\theta+4cos^4\theta)\nonumber\\
&-\frac{2\bar{g}^2}{\bar{\eta}} \frac{1}{1+\bar{m}^2}\int\mathrm{d\Omega_d}
(\cos^4\theta-\cos^2\theta)\nonumber\\
&+\frac{6\bar{g}^2\bar{\lambda}}{\bar{\eta}^2} \int\mathrm{d\Omega_d}
(-1+\cos^2\theta)\cos^2\theta\nonumber\\
&+\frac{2}{d+2} \frac{6\bar{g}^2\bar{\lambda}}{\bar{\eta}^2} (1+\bar{m}^2)\int\mathrm{d\Omega_d}
(-1+\cos^2\theta) \nonumber\\
&-\frac{3\bar{g}^2\bar{\lambda}}{\bar{\eta}^3} \int\mathrm{d\Omega_d}
(4\cos^2\theta-4\cos^4\theta) \nonumber\\
&+\frac{2}{d+4} \frac{32\bar{g}^2 \bar{\lambda}^2}{\bar{\eta}^3} (1+\bar{m}^2)\int\mathrm{d\Omega_d}
(\cos^2\theta-\cos^4\theta)\,, \label{eq:flow_lambda}\\
\partial_\tau \bar{\eta}=&-(z-2)\bar{\eta} \nonumber\\
&+\frac{2}{d} \frac{\bar{g}^2}{\bar{\lambda}} \frac{1}{(1+\bar{m}^2)^3}\int\mathrm{d\Omega_d}
(1-3\cos^2\theta+2\cos^4\theta)\,, \label{eq:flow_eta}\\
\partial_\tau \bar{g}=&-(z-3+(4-d)/2) \bar{g}\,, \label{eq:flow_g}
\end{align}
where $\mathrm{d\Omega_d}$ represents the angular integration in $d$-dimension. The renormalized dimensionless mass square of the order parameter reads
\begin{align}
\bar{m}^2=u'(\bar{\rho})+2\bar{\rho}u^{(2)}(\bar{\rho})\,.
\end{align}
As for the static anomalous dimension $\eta_\phi$, we utilize the result with the modified local potential approximation (LPA$^\prime$) 
\begin{align}
  \eta_\phi=&\frac{8}{d}\frac{1}{2^d\pi^{d/2}\Gamma(d/2)}\frac{\bar \rho_0 \big(u^{(2)}(\bar\rho_0)\big)^2}{\big(1+2\bar \rho_0 u^{(2)}(\bar{\rho}_0)\big)^2}\,,\label{eq:eta-LPApri}
\end{align}
where $\bar{\rho}_0$ is the minimum of the effective potential, i.e., $u'(\bar \rho_0)=0$.

%
\begin{figure*}[t]
\includegraphics[width=0.45\textwidth]{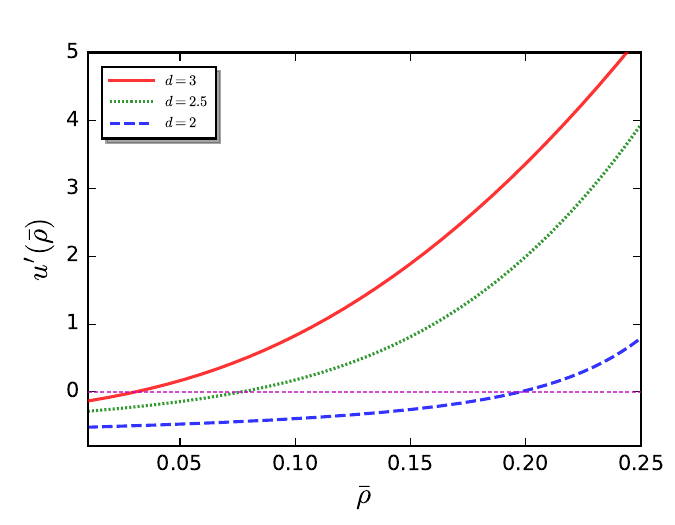}\hspace{0.5cm}
\includegraphics[width=0.45\textwidth]{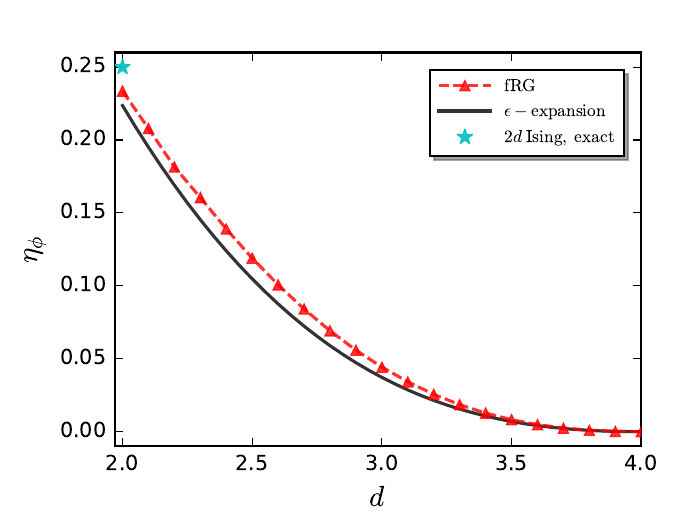}
\caption{Left panel: Global solution for $u'(\bar{\rho})$ derived from the fixed-point equation \Eq{eq:flow_potential}. Right panel: Static anomalous dimension $\eta_\phi$ as a function of the spatial dimension $d$, in comparison with the $\epsilon$-expansion result presented in \Eq{eq:eta-expansion}.}\label{fig:u-eta-d}
\end{figure*}
%

%
\begin{figure*}[t]
\includegraphics[width=0.45\textwidth]{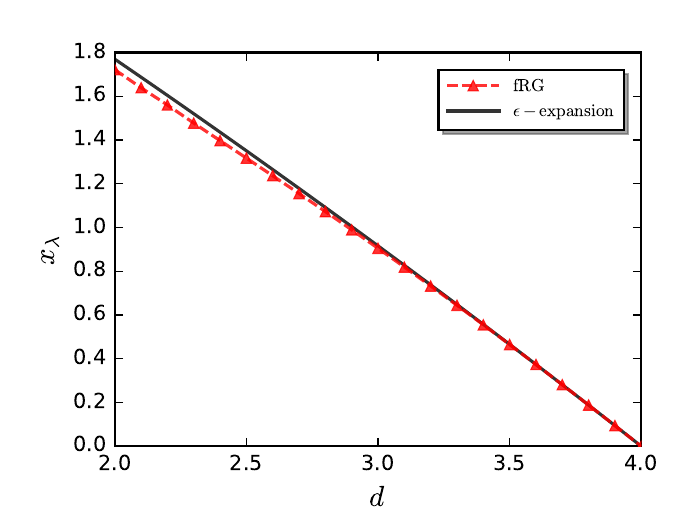}\hspace{0.5cm}
\includegraphics[width=0.45\textwidth]{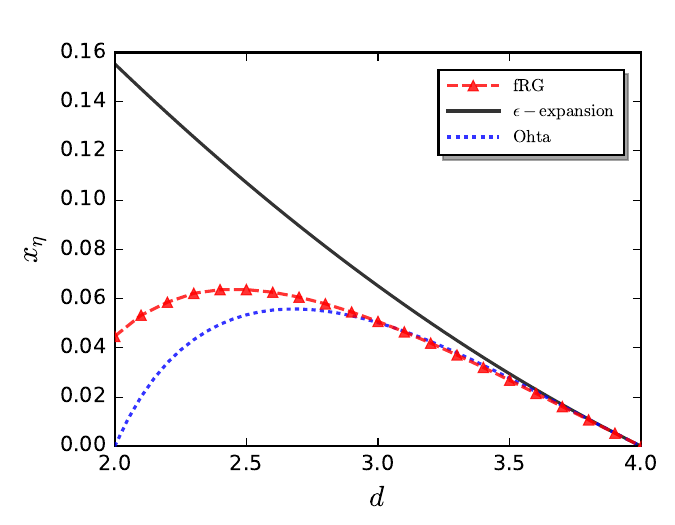}
\caption{Dynamic anomalous dimensions $x_\lambda$ (left panel) and $x_\eta$ (right panel) as functions of the spatial dimensions $d$. Results from the fRG and $\epsilon$-expansion in \cite{Siggia:1976zz} are compared. Moreover, the result of $x_\eta$ obtained by Ohta and Kawasaki in \cite{Ohta:1975} is also shown in the right panel in blue dashed line, see text for more details.}\label{fig:xlam-xeta-d}
\end{figure*}
%

Here we neglect the loop diagram correction for the coupling $g_k$. The flow for $\bar{g}$ only comes from the running of canonical dimension. It has been proved that there is no correction for the coupling to the one-loop order in the perturbative theory \cite{Siggia:1976zz, Yee:2017sir}. Besides, the relevant symmetries of the system also protect the coupling constant from renormalization. It has been shown that it is the combination of $\bar{\lambda}\bar{\eta}$ that determines the scaling behavior, rather than the coefficients $\bar{\lambda}$ and $\bar{\eta}$ separately. Therefore we define a new dimensionless coupling constant
\begin{align}
    f=\nu_d \frac{\bar{g}^2}{\bar{\lambda}\bar{\eta}}\,,
\end{align}
with $\nu_d=1/[2^d\pi^{d/2}\Gamma(d/2)]$ is the geometric factor corresponding to the angular integration. By combining the flow equations of \labelcref{eq:flow_lambda,eq:flow_eta,eq:flow_g}, one arrives at
\begin{align}
    \partial_\tau f=(\eta_\phi-(4-d))f-I_\lambda f^2-I_\eta f^2 \,,\label{eq:flow_f}
\end{align}
where the functions $I_\lambda, I_\eta$ read 
\begin{align}
I_\lambda=&-\frac{2}{(1+\bar{m}^2)^2}\int\mathrm{d\Omega_d} 
 (-1+\cos^2\theta) \cos^2\theta\nonumber\\
&+\frac{4}{d} \frac{1}{(1+\bar{m}^2)^2} \int\mathrm{d\Omega_d}
(2-6\cos^2\theta+4\cos^4\theta)\nonumber\\
&-\frac{2}{1+\bar{m}^2}\int\mathrm{d\Omega_d}
(\cos^4\theta-\cos^2\theta)\,,\label{eq:I-lam}\\
I_\eta=&\frac{2}{d} \frac{1}{(1+\bar{m}^2)^3}\int\mathrm{d\Omega_d}
(1-3\cos^2\theta+2\cos^4\theta)\,,
\end{align}
where the high-order terms in \Eq{eq:I-lam} from \Eq{eq:flow_lambda} have been neglected. The fixed-point equation of the coupling constant $f$ as well as its solution is readily obtained from \Eq{eq:flow_f}.

Once we find the fixed-point solution of $f^*$, the anomalous dimensions of transport coefficient $\lambda_k$ and shear viscosity $\eta_k$ can be easily obtained according to the scaling behavior in the critical regime
\begin{align}
\lambda_k\sim k^{-x_\lambda}, \quad \eta_k\sim k^{-x_\eta}\,,
\end{align}
where the exponents $x_\lambda$ and $x_\eta$ are the anomalous dimensions, and they can be computed using the following relation:
\begin{align}
x_\lambda=-I_\lambda f^*,\quad 
x_\eta=-I_\eta f^*\,.\label{eq:xlam_xeta}
\end{align}
Actually, \Eq{eq:xlam_xeta} can be obtained from the right hand side of flow equations in \Eq{eq:flow_lambda} and \Eq{eq:flow_eta}.
Finally, the dynamic critical exponent $z$ of the order parameter is obtained based on \Eq{eq:exponentz}. In the next section, we will show the numerical results of the fixed-point solution and various critical exponents.


\section{Numerical results}
\label{sec:num-resul}

%
\begin{figure}[t]
\includegraphics[width=0.45\textwidth]{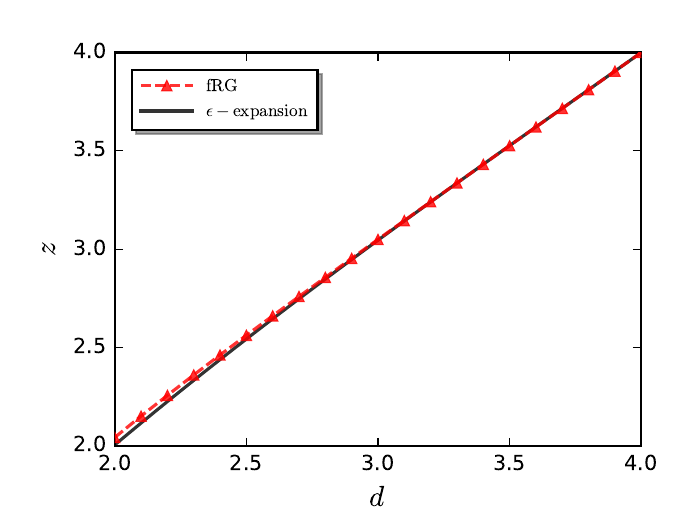}
\caption{Dynamical critical exponent $z$ defined in \Eq{eq:exponentz} as a function of the spatial dimension $d$. The $\epsilon$-expansion for $\eta_\phi$ in \Eq{eq:eta-expansion} and $x_\lambda$ in \Eq{eq:xlam-eta-expan} are also presented for comparison.}\label{fig:z-d}
\end{figure}
%

In this work, the derivative of global effective potential $u'(\bar{\rho})$ is obtained by using the large field expansion method proposed in \cite{Tan:2022ksv}. By directly solving the fixed-point equation starting from the large field asymptotic, we achieve the global effective potential with high numerical accuracy. The fixed-point value of the coupling constant $f^*$ is then determined from the global effective potential.

In the left panel of \Fig{fig:u-eta-d} we present the global solution of the first-order derivative of the effective potential $u'(\bar{\rho})$ as a function of $\bar{\rho}$ for three different spatial dimensions. The relevant static anomalous dimension $\eta_\phi$ in LPA$^\prime$ in \Eq{eq:eta-LPApri} is shown in the right panel of \Fig{fig:u-eta-d}. The function $u'(\bar{\rho})$ demonstrates monotonic behavior with respect to $\bar{\rho}$ and possesses a unique zero crossing point, indicating the presence of the Wilson-Fisher fixed point. As the spatial dimension $d$ decreases, the position of zero point $\bar{\rho}_0$ increases, indicating the necessity of a global solution rather than a local Taylor expansion, as discussed in \cite{Tan:2022ksv}. In the LPA$'$ approximation, the static anomalous dimension follows the form given in \Eq{eq:eta-LPApri}, and we compare our results with the $\epsilon$-expansion up to the third order \cite{Wilson:1971vs}.
\begin{align}
    \eta=\frac{N+2}{2(N+8)^2}\epsilon^2+\frac{N+2}{2(N+8)^2}\left[\frac{6(3N+14)}{(N+8)^2}-\frac{1}{4}\right]\epsilon^3\,. \label{eq:eta-expansion}
\end{align}
The value of $\eta_\phi$ from fRG computation and $\epsilon$-expansion are comparable when $d\geq3.5$. We also present the exact value $\eta_\phi=0.25$ for $d=2$ with a star in this plot.

In \Fig{fig:xlam-xeta-d}, we investigate the dependence of the dynamic anomalous dimensions $x_\lambda$, $x_\eta$ on the spatial dimension $d$. The $\epsilon$-expansion results from \Eq{eq:xlam-eta-expan} are also shown for comparison. The anomalous dimension $x_\lambda$, associated with the transport coefficient, shows excellent agreement with the perturbative expansion up to $\epsilon^2$, with deviations appearing for spatial dimensions $d<2.8$. In contrast, the behavior of $x_\eta$ differs. Results from the fRG calculations and $\epsilon$-expansion are comparable only as the spatial dimension approaches $d=4$. Additionally, our result for $x_\eta$ exhibits non-monotonic behavior when $d\lesssim2.5$. In \cite{Ohta:1975}, the exponent for $x_\eta$ is evaluated within the formulation of self-consistent mode coupling, where it is found 
\begin{align}
x_\eta =&-(2-\eta_\phi)^2\Gamma(1-\frac{\eta_\phi}{2})\Gamma(d-1-\frac{x_\eta}{2}+\frac{\eta_\phi}{2})\nonumber\\ &\times\Gamma(2+\frac{x_\eta}{2})\Big\{4(d-1)\Gamma(\frac{d}{2}+2)\Gamma(2-\frac{d}{2}+\frac{x_\eta}{2}-\frac{\eta_\phi}{2})\nonumber\\
&\times\Gamma(\frac{d}{2}-1-\frac{x_\eta}{2})\Gamma(\frac{d}{2}+\frac{\eta_\phi}{2})\Big\}^{-1}\,.\label{eq:xeta_itera}
\end{align}
The relevant result is shown in the right panel of \Fig{fig:xlam-xeta-d} in blue dashed line, obtained through numerical iterations for \Eq{eq:xeta_itera}, where the anomalous dimension $\eta_\phi$ are used from fRG calculation. Our result is comparable with this result as $d \gtrsim 3$.

Finally, the dynamic critical exponent $z$ as a function of spatial dimension $d$ is shown in \Fig{fig:z-d}. The anomalous dimensions $\eta_\phi$ and $x_\lambda$ are obtained from \Fig{fig:u-eta-d} and \Fig{fig:xlam-xeta-d} for our real-time fRG computation, respectively. The $\epsilon$-expansion results are taken from \Eq{eq:eta-expansion} for $\eta$ and from \Eq{eq:xlam-eta-expan} for $x_\lambda$. Since the order parameter is a conserved quantity, the dynamic critical exponent $z=4$ for spatial dimension $d=4$. Both results are highly consistent, with only minor differences emerging as the spatial dimension approaches $d=2$.

\section{Summary and outlook}
\label{sec:summary}
In this work, we investigate the critical dynamics of Model H, which describes a conserved order parameter field coupled with the transverse momentum density, using the real-time fRG approach. The Langevin equation of Model H incorporates kinetic terms and mode-coupling interactions. By integrating the noise terms and introducing the Martin-Siggia-Rose response field, i.e., the ``q'' type fields, which encode fluctuation effects in the Schwinger-Keldysh field theory, we obtain the RG scale-dependent effective action within the fRG approach. This formalism is well-suited for investigating the critical dynamics of Model H.

We solve the one-point and two-point correlation functions of the effective action to derive the flow equations for the effective potential, transport coefficient $\lambda_k$, and shear viscosity $\eta_k$. The contributions of mode couplings to the flow of $\lambda_k$ and $\eta_k$ are considered. We introduce the fixed-point equation for a coupling constant $f$, which determines the dynamical scaling behavior. The global fixed-point solution of the effective potential is obtained by integrating the flow equation from large to vanishing field. Subsequently, the anomalous dimensions for the transport coefficient $x_\lambda$ and shear viscosity $x_\eta$ are determined according to their scaling forms in the critical regime. Finally, the dynamic critical exponent for the order parameter is computed.

Our results for the anomalous dimension $x_\lambda$ and the dynamic critical exponent show excellent agreement with the perturbative $\epsilon$-expansion results in the range $2\leq d\leq 4$. Only small deviations emerge as the spatial dimension approaches $d=2$. However, result for anomalous dimension $x_\eta$ from fRG computation and $\epsilon$-expansion are comparable only in a narrow range of spatial dimensions close to $d=4$. Additionally, $x_\eta$ exhibits a non-monotonic behavior for $d\lesssim2.5$. From the analysis in \cite{Son:2004iv}, the dynamics of QCD critical end point and Model H belong to the same dynamic universality class. Our results predict a dynamic critical exponent $z=3.0507$ for $d=3$.

After this work is finished, we note that the critical dynamics of Model H is also studied within the real-time fRG independently in \cite{Roth:2024hcu}.

\begin{acknowledgments}
We thank Jan M. Pawlowski, Fabian Rennecke and Vladimir V. Skokov for insightful comments and discussions. This work is supported by the National Natural Science Foundation of China under Grant Nos. 12175030. 
\end{acknowledgments}

\appendix
\section{Propagators and mode-coupling vertices}
\label{app:pro_vertex}
As mentioned in \sec{sec:ModelH}, in the real-time fRG framework, the propagators corresponding to the inverse of the two-point functions of the effective action, contain the retarded, advanced and Keldysh components. The general definition of the propagator is
\begin{align}
G_{\Phi_i\Phi_j,k}=\left(\frac{\delta^2\Gamma_k[\Phi]}{\delta\Phi_i\delta\Phi_j}+R_{\Phi,k}\right)^{-1}\,.
\end{align}
Taking the functional derivative with respect to the different types of fields, one arrives at the specific forms of the propagators as follows
\begin{align}
&i G^{R}_{\phi,k}(q)
  =\frac{i}{q_0+i \lambda_k \bm{q}^2 \left( \bm{q}^2 \Big(1+r_B\Big(\frac{\bm{q}^2}{k^2}\Big)\Big)+ m_{\phi,k}^2 \right)}\,,\nonumber\\[2ex]
  &i G^{A}_{\phi,k}(q)
  =\frac{i}{q_0-i \lambda_k \bm{q}^2 \left( \bm{q}^2 \Big(1+r_B\Big(\frac{\bm{q}^2}{k^2}\Big)\Big)+ m_{\phi,k}^2 \right)}\,,\nonumber\\[2ex]
  &i G^{K}_{\phi,k}(q)
  =\frac{4 \lambda_k\bm{q}^2}{q_0^2+ \left[\lambda_k \bm{q}^2 \left( \bm{q}^2 \Big(1+r_B\Big(\frac{\bm{q}^2}{k^2}\Big)\Big)+ m_{\phi,k}^2 \right)\right]^2 }\,,\label{eq:order-propagator}
\end{align}
for the order parameter and
\begin{align}
&i G^{R}_{\bm{j}_{\alpha\beta},k}(q)
  =\frac{i}{q_0+i \eta_k  \bm{q}^2 \Big(1+r_B\Big(\frac{\bm{q}^2}{k^2}\Big)\Big)} \Pi^\bot_{\alpha\beta}\,,\nonumber\\[2ex]
  &i G^{A}_{\bm{j}_{\alpha\beta},k}(q)
  =\frac{i}{q_0-i \eta_k  \bm{q}^2 \Big(1+r_B\Big(\frac{\bm{q}^2}{k^2}\Big)\Big)} \Pi^\bot_{\alpha\beta},,\nonumber\\[2ex]
  &i G^{K}_{\bm{j}_{\alpha\beta},k}(q)
  =\frac{4 \eta_k\bm{q}^2}{q_0^2+ \left[\eta_k \bm{q}^2 \Big(1+r_B\Big(\frac{\bm{q}^2}{k^2}\Big)\Big)\right]^2}\Pi^\bot_{\alpha\beta} \,.\label{eq:j-propagator}
\end{align}
for the transverse momentum density fields. Since the effective potential satisfy the Ising symmetry, the order parameter only has the radial component. The mass term in \Eq{eq:order-propagator} reads
\begin{align}
m_{\phi,k}^2=V'_k(\rho_c) + 2\rho_c V''_k(\rho_c)\,.
\end{align}
The operator $\Pi^\bot$ in \Eq{eq:j-propagator} selects the transverse part of vector momentum density. The regulator functions are chosen as \cite{Litim:2000ci, Litim:2001up}
\begin{align}
R_{\phi,k}^{qc}(q)=i\lambda_k\bm{q}^4 r_B\Big(\frac{\bm{q}^2}{k^2}\Big)\,,\quad R_{j,k}^{qc}(q)=i\eta_k\bm{q}^2r_B\Big(\frac{\bm{q}^2}{k^2}\Big)\delta_{\mu\nu}\,,
\end{align}
where $r_B(x)$ is a Heaviside step function $\Theta(x)$, that is
\begin{align}
r_B(x)=(\frac{1}{x}-1)\Theta(1-x)\,.
\end{align}
For the flow equation of effective potential, the vertex $\Gamma^{(3)}_{\phi_q\phi_c\phi_c,k}$ emerging in the right hand of \Eq{eq:flowV} is
\begin{align}
i\Gamma^{(3)}_{\phi_q\phi_c\phi_c,k}=-3\rho_c^{1/2} V_k^{(2)}(\rho_c)-2\rho_c^{3/2}V_k^{(3)}(\rho_c)\,.
\end{align} 
To evaluate the flow equation for the two-point correlation function in \Fig{fig:Gam2-equ}, higher-order vertices are needed. They can be easily obtained from the functional derivative of $\Gamma_k[\Phi]$ w.r.t the corresponding fields $\phi$(or $j$). In this work, the three-point vertices arising from the coupling modes read
\begin{align}
&i\Gamma^{(3)}_{\phi_q\phi_c j_c,k}(q_1,q_2,q_3)=g_k \bm{q}_2\cdot\bm{e}_{j_c}\,,\\
&i\Gamma^{(3)}_{j_q\phi_c\phi_c,k}(q_1,q_2,q_3)=-g_k \bm{q}_3^2\bm{e}_{j_q}\cdot\bm{q}_2-g_k \bm{q}_2^2\bm{e}_{j_q}\cdot\bm{q}_3\,,
\end{align}
where the explicit momentum dependence from the ``classical'' fields are labelled. And the $\bm{e}_j$ denotes the unit vector along the $\bm{j}$ field direction.

\begin{widetext}
\section{Loop diagram calculation}
\label{app:loop}
In this appendix, we present a specific example to show how the calculation are performed and to illustrate the approximations used. We employ a loop diagram for the two-point correlation function $\Gamma_{\phi_q\phi_c,k}^{(2)}$ with the regulator function inserted on the propagator of order parameter, that is,
\begin{align}
\parbox[c]{0.15\textwidth}{\includegraphics[width=0.15\textwidth]{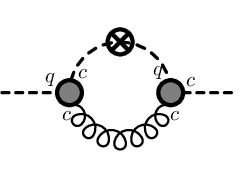}}=\int_q\partial_\tau R_{\phi,k} G_{\phi,k}^A(q) \Gamma^{(3)}_{\phi_q\phi_c j_c,k} G_{j,k}^K(q-p) \Gamma^{(3)}_{\phi_q\phi_c j_c,k} G_{\phi,k}^A(q)\,.
\end{align}
Substituting the expressions of vertices and propagators in \app{app:pro_vertex} into the equation above, one has
\begin{align}
&\int_q \partial_\tau R_{\phi,k}  \frac{1}{\left[iq_0+\lambda_k \bm{q}^2 \left( \bm{q}^2 \Big(1+r_B\Big(\frac{\bm{q}^2}{k^2}\Big)\Big)+ m_{\phi,k}^2 \right)\right]^2} \times
 \frac{4 \eta_k\bm{(q-p)}^2}{(q_0-p_0)^2+ \left[\eta_k \bm{(q-p)}^2 \Big(1+r_B\Big(\frac{\bm{(q-p)}^2}{k^2}\Big)\Big)\right]^2} \nonumber\\ 
 &\qquad g_k(i\bm{q}_\mu)\times\left(\delta^{\mu\nu}-\frac{\bm{(q-p)}_\mu \bm{(q-p)}_\nu}{\bm{(q-p)}^2}\right)\times g_k(-i\bm{p}_\nu)\nonumber\\
=&\int_q \partial_\tau R_{\phi,k} g_k^2(-1+\cos^2\theta)\bm{q}^2\bm{p}^2
\frac{4 \eta_k}{q_0^2+ \left[\eta_k \bm{q}^2 \Big(1+r_B\Big(\frac{\bm{q}^2}{k^2}\Big)\Big)\right]^2} \frac{1}{\left[iq_0+\lambda_k \bm{q}^2 \left( \bm{q}^2 \Big(1+r_B\Big(\frac{\bm{q}^2}{k^2}\Big)\Big)+ m_{\phi,k}^2 \right)\right]^2}\,.\label{eq:loopdiagram}
\end{align}
The vertices contribute to the external momentum of order $p^2$. Since the flow equation for transport coefficient $\lambda_k$ should be projected onto the $p^4$ term, we need to expand the $G_{j,k}^K$ propagator about the external momentum. The result is given by
\begin{align}
-\frac{1}{2} \frac{1}{[q_0^2/k^2+(\eta_kx(1+r_B(x)))^2]^2}2\eta_k^2\delta(1-x) 
\frac{4(\bm{q}\cdot\bm{p})^2}{k^4}\,,
\end{align}
where $x=\bm{q}^2/k^2$ is the ratio of internal momentum over RG scale.

Then we complete the integration in \Eq{eq:loopdiagram}, 
\begin{align}
\frac{1}{(2\pi)^d}\int \mathrm{d}x\mathrm{d\Omega_D} k^d x^{\frac{d}{2}} 16g_k^2 \lambda_k\eta_k^3 (-1+\cos^2\theta)\cos^2\theta \bm{q}^4\bm{p}^4\frac{1}{k^{16}} \frac{k^6(3\eta_k+\lambda_k k^2x(1+\bar{m}^2))}{4\eta_k^3(\eta_k+\lambda_k k^2x(1+\bar{m}^2))^3}\delta(1-x)\,.
\end{align}
As the RG scale evolves towards the infrared, the shear viscosity $\eta_k$ increases, and $\lambda_k$ approaches to a finite value. Then we have $\eta_k>>\lambda_k$, and the $\lambda_kk^2$ term in the denominator can be safely neglected, which leads us to
\begin{align}
\frac{1}{(2\pi)^d}\int \mathrm{d\Omega_D} \frac{6\bar{g}^2}{\bar{\eta}^2} (-1+\cos^2\theta)\cos^2\theta \lambda_k \bm{p}^4\,.
\end{align}
\end{widetext}
	
\bibliography{ref-lib}

\end{document}